\documentclass[prd,nofootinbib,floats,superscriptaddress,eqsecnum,tightenlines,11pt]{revtex4}

\usepackage{hyperref}
\usepackage{graphicx}
\usepackage{amsmath,amssymb,amsfonts,amsthm,latexsym,stmaryrd}
\usepackage{marginnote}
\usepackage{color}

\def\be{\begin{equation}}
\def\ee{\end{equation}}
\def\ba{\begin{eqnarray}}
\def\ea{\end{eqnarray}}
\def\bas{\begin{subequations}\begin{eqnarray}}
\def\eas{\end{eqnarray}\end{subequations}}

\def\lp{\ell_\text{Pl}}

\def\la{\langle}
\def\ra{\rangle}
\def\de{\mathrm{d}}

\def\f{\frac}

\def\SU{\text{SU}}

\def\SL{\text{SL}}
\def\su{\mathfrak{su}}

\def\sll{\mathfrak{sl}}

\def\L{{\text{\tiny{L}}}}
\def\E{{\text{\tiny{E}}}}
\def\lc{\ell_\text{c}}

\newtheorem{property}{Property}

\newtheorem{definition}{Definition}

\begin{document}

\title{BTZ Black Hole Entropy and the Turaev--Viro model}

\author{Marc Geiller}
\email{mgeiller@gravity.psu.edu}
\affiliation{Institute for Gravitation and the Cosmos \& Physics Department, Penn State, University Park, PA 16802, U.S.A.}
\author{Karim Noui}
\email{karim.noui@lmpt.univ-tours.fr}
\affiliation{Laboratoire de Math\'ematiques et Physique Th\'eorique, Universit\'e Fran\c cois Rabelais, Parc de Grandmont, 37200 Tours, France}
\affiliation{Laboratoire APC -- Astroparticule et Cosmologie, Universit\'e Paris Diderot Paris 7, 75013 Paris, France}

\begin{abstract}
We show the explicit agreement between the derivation of the Bekenstein--Hawking entropy of a Euclidean BTZ black hole from the point of view of spin foam models and canonical quantization. This is done by considering a graph observable (corresponding to the black hole horizon) in the Turaev--Viro state sum model, and then analytically continuing the resulting partition function to negative values of the cosmological constant.
\end{abstract}

\maketitle

%\tableofcontents

\section{Introduction}

\noindent In three spacetime dimensions, gravity is topological and has no local and only topological degrees of freedom. In spite of this fact, three-dimensional gravity with a negative cosmological constant admits black hole solutions known as the BTZ black holes \cite{BTZ,BHTZ}, which quite surprisingly exhibit thermodynamical properties and posses a Bekenstein--Hawking entropy. Although the origin of this entropy and the precise nature of the underlying microstates still remains somehow mysterious, an enormous amount of work has been devoted to the understanding of this puzzle, and numerous quantum gravity inspired state counting methods have been shown to yield the expected entropy formula (see \cite{Carlip} and references therein). The BTZ black hole has been a very fruitful toy model to test various ideas from quantum gravity in more than three spacetime dimensions, and its study has provided the first concrete example of the celebrated AdS/CFT correspondence \cite{Brown-Henneaux}. Its relevance is also much appreciated in string theory, where it turns out that the geometry of most near-extremal black holes can be described in terms of the BTZ solution (see \cite{Carlip2} and references therein). A natural question to ask is how the entropy of the BTZ black hole can be derived from the two complementary non-perturbative and background independent approaches to quantum gravity known as canonical loop quantum gravity (LQG hereafter) and spin foam models. Once such a description is available, one can further investigate which lessons it teaches for the description of the microstates of four-dimensional black holes in these two approaches, and finally the relationship with all the other proposals to derive the entropy.

The derivation of the BTZ black hole entropy from canonical LQG has been proposed only recently in \cite{FGNP-BTZ}, and in our opinion the derivations from state sum models proposed in \cite{SKG,GI1}, although clearly interesting, are not fully convincing. The reason for which the spin foam model derivations of \cite{SKG,GI1} dot not fully agree with the canonical formulation of \cite{FGNP-BTZ} is twofold. First, they require the fixing of an arbitrary parameter in the area spectrum in order to obtain the correct factor of 1/4 in the relation between the entropy and the area. Second, and most importantly, they are based on the Turaev--Viro model\footnote{In fact the two papers \cite{SKG,GI1} show that only the classical group limit of the Turaev--Viro model (i.e. the Ponzano--Regge model) is needed in order to compute the leading order contribution to the entropy.}, which corresponds to three-dimensional quantum gravity with a positive cosmological constant, but in this case there are in fact no black hole solutions, since three-dimensional black holes exist only in AdS. As we shall see in this work, these two issues are in fact related, and can be solved simultaneously. A first step towards this resolution was taken in \cite{GI2}, where an analytic continuation inspired by \cite{FGNP-BTZ} was proposed in the spin foam context in order to get back to the case of a negative cosmological constant. However, the calculation was carried out in such a way that the resulting entropy turns our to be proportional to $|k|^{-1}$, where $k$ is the (analytically-continued) Chern--Simons level. For these reasons, none of the existing proposals for the state sum model description of the entropy of a BTZ black hole can be considered as complete.

The technical difficulty in providing a quantum description of the BTZ black hole in the frameworks of LQG and spin foam models lies in the fact that these two approaches are under control and understood only when the underlying gauge group is compact, whereas the BTZ black hole involves non-compact gauge groups.  Indeed, this black hole is a solution of three-dimensional Lorentzian gravity in the presence of a negative cosmological constant, which is equivalent (upon invertibility of the triad field) to an $\SL(2,\mathbb{R})$ BF theory with a cosmological constant, or to an  $\SL(2,\mathbb{R})\times\SL(2,\mathbb{R})$ Chern--Simons theory.  Although in canonical LQG this non-compactness could potentially be dealt with, and the kinematical Hilbert space defined with an appropriate choice of regularization \cite{FL}, nothing is known about the physical states and the physical inner product (interesting developments have however recently appeared in \cite{DG} concerning the inclusion of a cosmological constant in LQG). In the spin foam approach the situation is even worse because the very definition of a state sum model for non-compact groups is only formal and involves many divergences that are not under control. In order to provide a description of the quantum Euclidean BTZ black hole, one could think of starting with a state sum model defined in terms of $\SL(2,\mathbb{C})$ quantum $6j$ symbols \cite{q6j1,q6j2,q6j3}. However, this task seems rather difficult since such models have so far not been investigated from the physical point of view, and it is not clear how exactly they relate to quantum gravity. We shall comment on this issue in the last section of this paper.

In spite of the above-mentioned difficulties, the framework of \cite{FGNP-BTZ} suggests a way of circumventing the problem of the non-compactness. The basic idea consists in looking for situations with compact symmetry groups, and then performing an analytic continuation in order to return to the case of physical interest. To do so, \cite{FGNP-BTZ} starts with two crucial steps. First, the Lorentzian signature is traded for the Euclidean one by a Wick rotation, and then the cosmological constant is taken to be positive. Therefore, one ends up working in the context of three-dimensional Euclidean gravity with a positive cosmological constant, for which the associated Chern--Simons theory is defined with the compact gauge group $\SU(2)\times\SU(2)$ and has a known quantization. In this context, one can compute physical observables $\mathcal{O}_\E(\Lambda>0)$, and then, after performing a suitable analytic continuation, one can define the corresponding physical observables $\mathcal{O}_\E(\Lambda<0)$ in Euclidean quantum gravity with a negative cosmological constant. By defining an observable associated with the horizon of a Euclidean BTZ black hole, one can obtain an explicit formula for the number of its microstates, and show that its logarithm reproduces the Bekenstein--Hawking law in the semiclassical limit. Although this strategy is not fully satisfactory because it does not lead to a precise description of the underlying black hole microstates (whose origin still remains unknown), it allows to compute their number (at least in the Euclidean regime), and it is so far the only known way of doing so in three-dimensional LQG. One should also emphasize that most of the techniques used to compute the BTZ black hole entropy rely on similar ideas \cite{Carlip}. Therefore, the question of whether this method can be applied to spin foam models is quite natural, and a first attempt at doing so appeared soon in \cite{GI2} after the publication of the work \cite{FGNP-BTZ}. However, even though the idea of \cite{GI2} is very appealing in light of the previous discussion, the construction carried out in this paper is incomplete and does not lead to the correct proportionality coefficient of 1/4 between the entropy and the area. The purpose of the present paper is to revisit this construction, and to show that a proper computation of the discretized BTZ partition function \textit{\`a la} spin foam leads indeed to the correct semiclassical entropy formula. 

Our goal in this work is to compute the discretized partition function for the Euclidean BTZ black hole using the idea of analytic continuation presented above. To do so, one natural starting point is to consider the Turaev--Viro state sum model, which is known to provide a representation of the path integral for three-dimensional Euclidean gravity with a positive cosmological constant. The model is then written on a solid torus, since this corresponds to the topology of the Euclidean BTZ black hole. The horizon being an $\mathbb{S}_1$ circle at the core of the torus, one can therefore define the observable partition function by fixing the spins coloring the edges of the horizon. The partition function obtained in this way is then a function of the spins $\vec{\jmath}=(j_1,\dots,j_n)$ coloring the $n$ edges of the discretized horizon. Each edge $e$ tessellating the horizon is associated with a quantum of length $L_e=8\pi\lp\sqrt{j_e(j_e+1)}$, where $\lp=G\hbar$ is the three-dimensional Planck length, and these microscopic contributions add up to give the macroscopic length of the horizon. By following this procedure, the Turaev--Viro state sum model in the presence of the horizon can be put in a very simple form, and we show that it reproduces exactly the dimension of the intertwiner space between the spins  $j_e$, $e\in\llbracket1,n\rrbracket$, viewed as representations of $\text{U}_q(\su(2))$, where the quantum deformation parameter $q$ is related to the cosmological constant. By performing the analytic continuation of \cite{FGNP-BTZ}, one finally obtains the number of BTZ black hole microstates, the logarithm of which reproduces the semiclassical Bekenstein--Hawking relation. Furthermore, the boundary states defined by the spin foam model in the presence of the black hole observable correspond exactly to the physical quantum states of the canonical theory introduced in \cite{FGNP-BTZ}. This establishes the connection between the canonical and covariant quantization schemes.

This paper is organized as follows. We start in section \ref{sec2} by recalling some basic properties of the Euclidean BTZ black hole and its topology. Section \ref{sec3} is devoted to reviewing the Turaev--Viro partition function in the presence and absence of boundaries, and the notion of state sum observables. In section \ref{sec4} we compute the partition function (for $\Lambda>0$) for the black hole observable, and shows that it satisfies a recursion relation that can be written in a closed form. We then show that once the analytic continuation is performed, the partition function reproduces exactly the number of microstates obtained in \cite{FGNP-BTZ} in the context of canonical quantization. Section \ref{sec5} contains some remarks concerning the Barbero--Immirzi parameter of LQG, the logarithmic corrections, and potential future developments. Finally, we summarize our result and present our conclusions in section \ref{sec6}.

\section{Geometry and topology of the BTZ black hole}
\label{sec2}

\noindent Before studying the quantum theory, we first recall some basic features of the BTZ solution. We will focus on the Euclidean BTZ black hole, which is obtained from the Lorentzian solution by a Wick rotation. After describing the geometry and topology of the Euclidean BTZ black hole, we then propose different graphical representations of this spacetime that will be useful when writing down the Turaev--Viro model and making the contact with the canonical theory.

In three spacetime dimensions, all the solutions to Einstein's equations are necessarily of constant curvature, and therefore look locally like homogeneous spaces. For instance, Lorentzian solutions with a negative cosmological constant are locally anti-de Sitter ($\text{AdS}_3$), whereas Euclidean solutions with a positive cosmological constant are locally spherical ($\mathbb{S}_3$). These are the two cases of interest for the purpose of this paper. Homogeneous spaces $\mathcal{H}$ are therefore the maximally symmetric solutions in three-dimensional gravity, and the corresponding isometry group $G$ acts faithfully on them. Any other solution $\mathcal{M}$ can be obtained as a (right or left) coset $\mathcal{M}=\mathcal{H}/H$ of $\mathcal{H}$ by a discrete subgroup $H$ of the isometry group $G$. For example, the Lorentzian BTZ black hole spacetime can be constructed as a quotient of $\text{AdS}_3$ by a discrete subgroup of $\SL(2,\mathbb{R})\times\SL(2,\mathbb{R})$ \cite{BHTZ}, which is totally defined by two parameters corresponding physically to the mass and the angular momentum of the black hole. Despite the apparent simplicity of the BTZ solution, and even though it has no curvature singularity at its center $r=0$, it shares many features with the four-dimensional Kerr solution. In particular it admits an event horizon, it possesses a Hawking temperature, and an entropy given by $S=L/(4\lp)$, where $L$ is the perimeter of the horizon. It is the existence of these thermodynamical properties that make the study of the BTZ black hole particularly interesting for the understanding of aspects of quantum black hole physics in four spacetime dimensions.

Let us now describe the BTZ solution in more details. As already mentioned, the BTZ black hole is a solution of three-dimensional Lorentzian pure gravity with a negative cosmological constant $\Lambda=-1/\lc^2$. For a convenient choice of Schwarzschild-type coordinates $(t,r,\phi)$, its metric is given by \cite{BTZ}
\be\label{lorentzian metric}
\de s_\L^2=-N^2\de t^2+N^{-2}\de r^2+r^2\big(\de\phi+N_\L^\phi\de t\big)^2,
\ee
where the lapse and shift functions have the following form:
\be
N(r)=\left(-8GM_\L+\f{r^2}{\lc^2}+\f{16G^2J_\L^2}{r^2}\right)^{1/2},\qquad N_\L^\phi(r)=-\f{4GJ_\L}{r^2},
\ee
and where the subscript {\tiny{L}} is here to indicate the Lorentzian quantities. $G$ denotes the three-dimensional Newton constant, and the conserved charges $M_\L$ and $J_\L$ are respectively the mass and angular momentum of the black hole. The outer event horizon and inner Cauchy horizon (which exists only when the angular momentum $J_\L$ is non-vanishing) are defined by the expressions
\be
r^2_\pm=4GM_\L\lc^2\left(1\pm\sqrt{1-\left(\f{J_\L}{M_\L\lc}\right)^2}\right),
\ee
which implies necessarily that $M_\L>0$ and $|J_\L|\leq M_\L\lc$. The BTZ black hole being locally of constant negative curvature, it is isometric to the three-dimensional anti-de Sitter spacetime $\text{AdS}_3$. Globally, it is defined as the coset $\text{AdS}_3/H$ of $\text{AdS}_3$ by a discrete subgroup $H$ of 
the homogeneous space's isometry group $\SL(2,\mathbb{R})\times\SL(2,\mathbb{R})$, and the definition of $H$ depends on the mass $M_\L$ and the angular momentum $J_\L$ of the black hole.

Interestingly, there exists also a BTZ black hole solution in the case of a Euclidean signature \cite{EuclideanBTZ}. Its metric can be obtained from \eqref{lorentzian metric} by analytically continuing $t\rightarrow-i\tau$, $M_\L\rightarrow M$, and $J_\L\rightarrow iJ$. The corresponding solution is then given by
\be\label{euclidean metric}
\de s^2=(N^\perp)^2\de\tau^2+f^{-2}\de r^2+r^2\big(\de\phi+N^\phi\de\tau\big)^2,
\ee
with
\be
N^\perp(r)=f(r)=\left(-8GM+\f{r^2}{\lc^2}-\f{16G^2J^2}{r^2}\right)^{1/2},\qquad N^\phi(r)=-iN_\L^\phi(r)=-\f{4GJ}{r^2},
\ee
and the horizons are located at the distances $r_\pm$ given by the equations:
\be\label{euclidean radius}
r^2_\pm=4GM\lc^2\left(1\pm\sqrt{1+\left(\f{J}{M\lc}\right)^2}\right).
\ee
Just like its Lorentzian counterpart, the Euclidean BTZ black hole solution is locally of constant negative curvature. However, due to the signature change, it is isometric to the hyperbolic three-space $\mathbb{H}_3$. Globally, it can be obtained as the coset $\mathbb{H}_3/H$ of $\mathbb{H}_3$ by a discrete subgroup of the three-dimensional (universal covering of the) Lorentz group $\SL(2,\mathbb{C})$, this latter being the isometry group of $\mathbb{H}_3$. The resulting topology is that of a solid torus, with the horizon corresponding to a circle at the core of this torus. In spite of these facts being well-known and having been established quite some time ago, it is worth explaining in more details how one obtains the topology of a solid torus from the Euclidean BTZ solution, since this observation will be crucial for our construction in section \ref{sec4}.

To exhibit the topology of the solid torus, let us first consider the coordinate transformation \cite{EuclideanBTZ}
\begin{subequations}
\ba
x&=&\sqrt{\f{r^2-r_+^2}{r^2-r_-^2}}\cos\left(\f{r_+}{\lc^2}\tau+\f{|r_-|}{\lc}\phi\right)\exp\left(\f{r_+}{\lc}\phi-\f{|r_-|}{\lc^2}\tau\right),\\
y&=&\sqrt{\f{r^2-r_+^2}{r^2-r_-^2}}\sin\left(\f{r_+}{\lc^2}\tau+\f{|r_-|}{\lc}\phi\right)\exp\left(\f{r_+}{\lc}\phi-\f{|r_-|}{\lc^2}\tau\right),\\
z&=&\sqrt{\f{r_+^2-r_-^2}{r^2-r_-^2}}\exp\left(\f{r_+}{\lc}\phi-\f{|r_-|}{\lc^2}\tau\right) \, >0,
\ea
\end{subequations}
which takes the metric to the standard Poincar\'e form on the upper half-space ($z>0$):
\be
\de s^2=\f{\lc^2}{z^2}\big(\de x^2+\de y^2+\de z^2\big),
\ee
provided we set  $|r_-|=ir_-={4JG\lc}/{r_+}$ since $r_-$ is purely imaginary as can be see from \eqref{euclidean radius}. One can further use the change of coordinates $(x,y,z)=(R\cos\theta\cos\chi,R\sin\theta\cos\chi,R\sin\chi)$ to spherical coordinates $(R,\theta,\chi)$, which brings the metric to the form
\be
\de s^2=\f{\lc^2}{\sin^2\chi}\left(\f{\de R^2}{R^2}+\cos^2\chi\de\theta^2+\de\chi^2\right).
\ee
Then, in order to account for the periodicity of the angular Schwarzschild coordinate $\phi$, one must proceed to the global identifications
\be \label{parameters}
(R,\theta,\chi)\sim\left(R\exp\left(\f{2\pi r_+}{\lc}\right),\theta+\f{2\pi|r_-|}{\lc},\chi\right).
\ee
A fundamental region for these identifications is simply the space between the hemispheres located at $R=1$ and $R=\exp(2\pi r_+/\lc)$, with boundary points identified along the radial lines and after a twist of angle $2\pi|r_-|/\lc$ (which is vanishing if there is no angular momentum $J$). Finally, we arrive at the well-known result that the topology of the Euclidean BTZ black hole is that of a solid torus $\mathbf{T}=\mathbb{D}_2\times\mathbb{S}_1$, with the three dimensions labelled by $R$, $\theta$ and $\chi$ being compactified.

\begin{center}
\begin{figure}[h]
\includegraphics[scale=0.5]{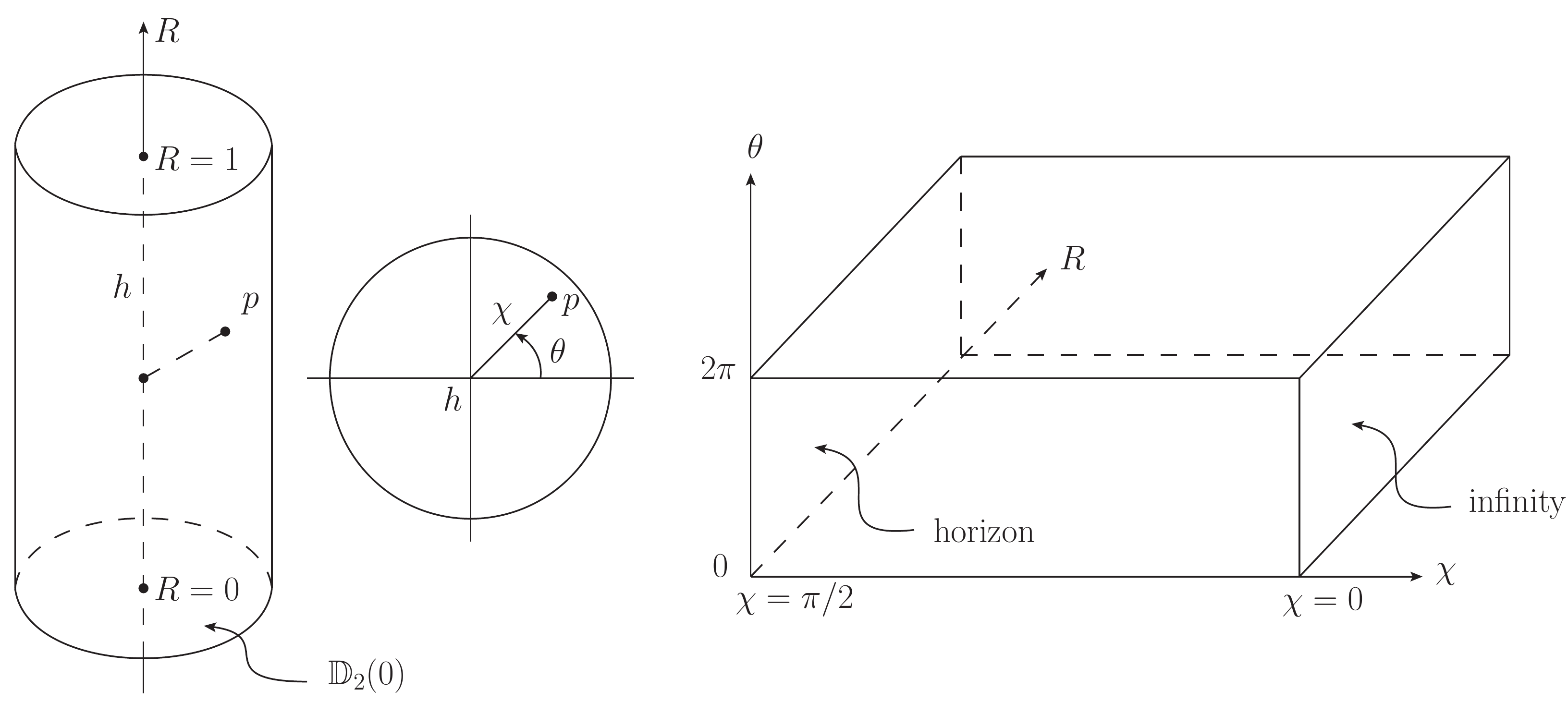}
\caption{Different representations of the Euclidean BTZ spacetime. On the left-hand side, the solid torus is represented as $\mathbb{D}_2\times[0,1]$, where $\mathbb{D}_2$ is the disk. The two discs $\mathbb{D}_2(0)$ and $\mathbb{D}_2(1)$ at the extremal values of the radial coordinate $R$ are identified. In this representation, the position of a point $p\in\mathbf{T}$ in the solid torus is determined by the coordinates $(R,\theta,\chi)$, where the $(\theta,\chi)$ slice at $R=\text{constant}$ is represented in the middle. On the right-hand side, the solid torus is represented by a parallelepiped with identifications. Any point $(R,\theta)$ of the surface $\chi=\pi/2$ is identified with a point $(R,0)$ on the horizon, and the planes $\theta=0$ and $\theta=2\pi$ are identified. Such a representation is particularly interesting if one wants to ``slice'' the BTZ spacetime in order to study the canonical theory.}
\label{BTZrepresentation}
\end{figure}
\end{center}

To represent graphically the Euclidean BTZ spacetime, it is simpler to focus on the non-rotating case $J=0$. In this case, the spacetime is still a solid torus, and the parameters \eqref{parameters} can be chosen such that $\chi\in[0,\pi/2]$, $\theta\in [0,2\pi]$, and $R\in[1,\exp(2\pi r_+/\lc)]$. From the previous formulae, it is immediate to see that the boundary of the spacetime is located at $\chi=0$, and therefore has the topology of a two-torus, namely $\mathbb{T}_2=\mathbb{S}_1\times\mathbb{S}_1$. Concerning the horizon, it  is located at the ``surface'' $\chi=\pi/2$, which degenerates into the one-dimensional manifold $\mathbb{S}_1$.  To make the contact with the canonical approach, we will need to identify (Euclidean) time slices in this set of variables, which can be done straightforwardly because the $\tau=\text{constant}$ surfaces are sent to $\theta=\text{constant}$ surfaces. In this sense, the angle $\theta$ represents the time variable. All these properties are summarized on figure \ref{BTZrepresentation}.

\section{The Turaev--Viro state sum model}
\label{sec3}

\noindent The Turaev--Viro model \cite{TV} is an assignment of a state sum to any compact (but not necessarily closed) oriented three-manifold. From the mathematical point of view, this state sum defines a topological invariant of the manifold (at least when this latter has no boundaries), or a knot invariant in the case where observables are included. From the physical point of view, the Turaev--Viro model gives an exact expression for the partition function of Euclidean three dimensional (first order) gravity in the presence of a positive cosmological constant.

The model is based on the representation theory of the quantum group $\text{U}_q(\su(2))$, where the quantum deformation parameter $q$ is a root of unity. The fact that $q$ is a root of unity is crucial because, as we shall see later on, it makes the state sum model mathematically well-defined. From the physical perspective, this condition on $q$ is a consequence of the gauge-invariance of the path integral measure for gravity. One has the following relationship between $q$, the three-dimensional Newton constant $G$, and the cosmological constant $\Lambda$:
\be\label{q and k}
q=\exp\left(\f{i\pi}{k+2}\right),\qquad k=\f{\lc}{\lp}=\f{1}{G\hbar\sqrt{\Lambda}}.
\ee
Here $k$ is the coupling constant (the level) of the Chern--Simons formulation of gravity. One can show that gauge-invariance requires that $k$ be an integer \cite{DJT}, which in turn implies that $q$ is indeed a root of unity (of order $2(k+2)$).

Contrary to its classical counterpart, $\text{U}_q(\su(2))$ admits only a finite number of unitary irreducible representations, which are labelled by a half-integer spin $j\leq k/2$. These (non-cyclic) representations are finite-dimensional, of dimension $d_j=2j+1$. It is useful to introduce the quantum dimension $\big[d_j\big]_q$ (obtained as the evaluation of the $\mu$-ribbon element in the spin $j$ representation), where the so-called quantum numbers are defined by the parameter $q$ according to
\be
[n]_q=\f{q^n-q^{-n}}{q-q^{-1}}=\f{\displaystyle\sin\left(\f{\pi}{k+2}n\right)}{\displaystyle\sin\left(\f{\pi}{k+2}\right)},\qquad\forall\,n\in\mathbb{N}-\{0\}.
\ee
With these quantities, we can now assign to any spin $j$ a weight
\be\label{edge weight}
\omega_j\equiv(\sqrt{-1})^{2j}\big[d_j\big]_q^{1/2} ,
\ee
and further introduce the positive real number $\omega$ defined by the relation
\be
\omega^2\equiv\sum_{j=0}^{k/2}\omega_j^4=\sum_{j=0}^{k/2}\big[d_j\big]_q^2=-\f{2(k+2)}{(q-q^{-1})^2}=\f{k+2}{2}\sin^{-2}\left(\f{\pi}{k+2}\right).
\ee
We are now ready to define the Turaev--Viro model. Its definition does actually depend on whether or not the compact manifold possesses a boundary. While the case of a closed manifold $M$ (i.e. when $\partial M=\varnothing$) is unambiguous, there is no unique way of defining the model in the presence of a boundary. For this reason, and because the treatment of the boundary might be an important issue in the understanding of black hole entropy (and it is indeed paramount in the continuum theory \cite{Brown-Henneaux}), we choose to treat both cases separately in the following two subsections.

\subsection{Manifold without boundary}

\noindent Let $M$ be a closed three-dimensional manifold, and $\Delta$ one of its triangulations consisting of zero-simplices $v$ (vertices), one-simplices $e$ (edges), two-simplices $t$ (triangles), and three-simplices $\tau$ (tetrahedra). Given this triangulation, one can define the amplitude
\be\label{TVsummand}
\mathcal{Z}(M,\Delta)=\prod_v\omega^{-2}\prod_e\omega_{j_e}^2\prod_\tau\big|6j_{e\subset\tau}\big|_q,
\ee
which is constructed by assigning to the vertices $v\in\Delta$ the weights $\omega^{-2}$, to the edges $e\in\Delta$ the weights $\omega^2_{j_e}$ where $j_e$ is a representation of $\text{U}_q(\su(2))$, and to the tetrahedra $\tau\in\Delta$ the quantum $6j$ symbols $\big|6j_{e\subset\tau}\big|_q$ defined\footnote{Notice that our definition, in agreement with \cite{TV}, absorbs the weights assigned to the triangles $t\in\Delta$ bounding the tetrahedra.} in appendix \ref{appendix:6jq}. As suggested by the notation, for each tetrahedron the quantum $6j$ symbol is a function of the six representations $j_e$ coloring the edges $e\subset\tau$ bounding the tetrahedron. Finally, given the quantity \eqref{TVsummand} which is obviously triangulation-dependent, the Turaev--Viro partition function is given by the following state sum:
\be\label{TVpartitionfunction}
\mathcal{Z}(M)=\sum_\phi\mathcal{Z}(M,\Delta).
\ee
The sum in this formula is taken over all the admissible labelings $\phi:\{e\in\Delta\}\rightarrow\{0,1/2,1,\dots,k/2\}$ of the edges $e\in\Delta$ by spins $j_e$ viewed as unitary irreducible representations of $\text{U}_q(\su(2))$. A labeling is said to be admissible if for any two-simplex (triangle) $t\in\Delta$, the three boundary edges $e\subset t$ are labeled by admissible spins, this admissibility condition on a triple $(i,j,m)$ of spins being given by\footnote{The $k$ appearing in the fourth condition is the Chern--Simons level defined in \eqref{q and k}, and not a spin label.}
\be\label{admin}
i\leq j+m,\qquad j\leq i+m,\qquad m\leq i+j,\qquad i+j+m\leq k,\qquad i+j+m=\hspace{-0.35cm}\mod1.
\ee
Note that these admissibility conditions are implicitly contained in the definition of the quantum $6j$ symbols, as can be seen in appendix \ref{appendix:6jq}.

An important property of the Turaev--Viro state sum $\mathcal{Z}(M)$ is that it is finite and, as indicated by the notation, that it does not depend on the choice of triangulation $\Delta$. It is a topological invariant that depends only on the topology of the manifold $M$. Its relation to the quantization of three-dimensional gravity can be understood in two ways. First of all, since the action for first order gravity can be written as two copies of Chern--Simons actions with opposite levels $k$ and $-k$ one has the well-known result that
\be\label{WRT}
\mathcal{Z}(M)=|\mathcal{Z}_\text{WRT}(M)|^2,
\ee
where $\mathcal{Z}_\text{WRT}(M)$ denotes the so-called Witten--Reshetikhin--Turaev canonical evaluation of the path integral (see \cite{AGN} for a review, and \cite{TVir} for a proof of the above relation). The second argument comes from a result of Mizoguchi and Tada, who showed that the asymptotic (semiclassical) behavior of the Turaev--Viro model was related to Regge gravity with a cosmological constant \cite{MT}. This is consistent with the fact that the classical group limit $q\rightarrow1$ of the Turaev--Viro model corresponds to the Ponzano--Regge model \cite{PR}, whose link with first-order gravity with $\Lambda=0$ is well understood (see \cite{PR1,PR2,PR3,PR4}).

\subsection{Manifold with boundary}
\label{sec:IIIb}

\noindent Let us now turn to the case where the manifold $M$ has a boundary. In this case, there are essentially two ways of extending the definition of the Turaev--Viro model. The first one, which is already described in the original paper \cite{TV}, consists in fixing once and for all the boundary triangulation and its coloring and to modify the weights accordingly, in such a way that the partition function is well behaved under cobordisms. The second modification consists in changing the type of boundary data in order for the resulting partition function be invariant also under boundary Pachner moves \cite{KMS,CCM}.

\subsubsection{Fixed boundary triangulation}

\noindent In full generality, a generic triangulation $\Delta$ of a manifold $M$ with a boundary can have edges and vertices lying in the boundary. Let us therefore divide the set of edges $e\in\Delta$ and of vertices $v\in\Delta$ into subsets consisting of boundary edges $e_b$, boundary vertices $v_b$, internal edges $e_i$, and internal vertices $v_i$. In other words, denoting the triangulation of the boundary $\partial M$ by $\partial\Delta$, we have that $e_b,v_b\in\partial\Delta$, and $e_i,v_i\in\Delta\backslash\partial\Delta$. With this distinction between boundary and internal contributions, we can introduce the quantity\footnote{Our notation is a bit redundant. Indeed, since a given triangulation $\Delta$ of $M$ does already define a unique boundary triangulation $\partial\Delta$ of $\partial M$, one could simply write $\mathcal{Z}(M,\Delta)$. However, we choose to write $\mathcal{Z}(M,\Delta,\partial\Delta)$ in order to highlight the difference between \eqref{TV summand with boundary} and \eqref{TVsummand}.}
\be\label{TV summand with boundary}
\mathcal{Z}(M,\Delta,\partial\Delta)=\prod_{v_b}\omega^{-1}\prod_{v_i}\omega^{-2}\prod_{e_b}\omega_{j_e}\prod_{e_i}\omega_{j_e}^2\prod_\tau\big|6j_{e\subset\tau}\big|_q.
\ee
Let us now denote by $\phi_b$ an admissible coloring of the boundary edges. By this, we mean a coloring of the edges $e_b$ with spins such that each triangle $t\in\partial\Delta$ has its three edges labelled by an admissible triple. Given such a fixed boundary data, one can then define the partition function
\be\label{TVwithbound}
\mathcal{Z}(M,\phi_b)=\sum_{\phi_i}\mathcal{Z}(M,\Delta,\partial\Delta),
\ee
where the sum is taken over all the admissible labelings $\phi_i$ of the internal edges $e_i$ compatible with the fixed coloring $\phi_b$ of the boundary triangulation. Clearly, the partition function \eqref{TVwithbound} is not fully triangulation independent, in the sense that it depends on the choice of triangulation and spin labelling of the boundary. However, one can show that it does not depend on the triangulation $\Delta\backslash\partial\Delta$ of the interior of the manifold \cite{TV}. In other words, it does not depend on the details of the triangulation away from the boundary.

The advantage of introducing \eqref{TV summand with boundary} is that it enables to encode the functorial nature of the Turaev--Viro invariant, and to compute the invariant of ``complicated'' manifolds by the technique of surgery. Indeed, if one decomposes a spacetime manifold $M$ without boundary into two pieces $M_1$ and $M_2$ that share the same boundary $\Sigma\equiv\partial M_1=\partial M_2$, whose triangulation is denoted by $\Delta_\Sigma$, one has the relation
\be
\mathcal{Z}(M,\Delta)=\mathcal{Z}(M_1,\Delta_1,\Delta_\Sigma)\mathcal{Z}(M_2,\Delta_2,\Delta_\Sigma),
\ee
and the invariant of $M$ can be obtained as
\be
\mathcal{Z}(M)=\sum_{\phi_b}\sum_{\phi_1}\sum_{\phi_2}\mathcal{Z}(M_1,\Delta_1,\Delta_\Sigma)\mathcal{Z}(M_2,\Delta_2,\Delta_\Sigma)=\sum_{\phi_b}\mathcal{Z}(M_1,\phi_b)\mathcal{Z}(M_2,\phi_b),
\ee
where $\phi_1$ (resp. $\phi_2$) denotes the admissible colorings of $\Delta_1\backslash\Delta_\Sigma$ (resp. $\Delta_2\backslash\Delta_\Sigma$), and $\phi_b$ the admissible colorings of the triangulation of $\Sigma$. For example, this can be used to obtain the Turaev--Viro invariant for a three-sphere by gluing together two three-balls along their common boundary (which is a two-sphere). This property of topological quantum field theories is the main reason for introducing the version \eqref{TVwithbound} of the Turaev--Viro model for a manifold with boundary.

\subsubsection{PL-homeomorphism invariance of the boundary}
\label{subsubsec:PL}

\noindent We have just seen that compared to the case of a manifold without boundary, the partition function \eqref{TVwithbound} satisfies a weaker form of triangulation-independence, in the sense that it is only independent of the choice of triangulation away from the boundary. However, one might also want to consider a model having the additional property of being invariant under elementary boundary operations, thereby defining a PL (piecewise linear)-homeomorphism invariant of both the manifold interior and its boundary. As shown in \cite{CCM} (see also \cite{KMS} for a different proceedure), this can easily be achieved by considering a slight modification of the quantity \eqref{TVsummand}, which consists in assigning a $q$-deformed Wigner $3jm$ symbol to the two-simplices (triangles) $t_b$ that lie in the boundary triangulation $\partial\Delta$. With this additional input, the amplitude of interest takes the form
\be\label{CCM summand}
\widetilde{\mathcal{Z}}(M,\Delta,\partial\Delta)=\prod_v\omega^{-2}\prod_e\omega_{j_e}^2\prod_{t_b}\big|3j_{e_b\subset t_b}m\big|_q\prod_\tau\big|6j_{e\subset\tau}\big|_q,
\ee
where
\be\label{triangle weight}
\big|3j_{e_b\subset t_b}m\big|_q\equiv
(-1)^{(m_1+m_2+m_3)/2}
\left(
\begin{array}{ccc}
j_1&j_2&j_3\\
m_1&m_2&-m_3
\end{array}\right)_q,
\ee
and the symbol between parenthesis is the quantum $3jm$ symbol (whose relation with the quantum Clebsch--Gordan coefficient is recalled in appendix \ref{appendix:6jq}). The invariant is then defined by taking the sum
\be
\widetilde{\mathcal{Z}}(M)=\sum_{\phi}\sum_{-j_{e_b}\leq m\leq j_{e_b}}\widetilde{\mathcal{Z}}(M,\Delta,\partial\Delta),
\ee
where $\phi$ are the admissible labelings of all possible edges (both internal and boundary), and for each spin $j_{e_b}$ assigned to a boundary edge there is an additional sum over the associated magnetic number $m$.

In \cite{OL} it was explained (in the classical group limit $q\rightarrow1$) that the Wigner $3jm$ symbols assigned to the boundary triangles arise naturally when one considers the construction of the Ponzano--Regge model from the first order action on a manifold with boundary, and the alternative choices corresponding to fixing either the boundary metric or the boundary connection were discussed. It would be interesting to investigate further the type of discrete two-dimensional theory that is induced on the boundary in the case of quantum groups, and the possible relationship with conformal field theory. Interestingly, the two-dimensional topological models obtained in \cite{BK} correspond to the boundary components of the partition function \eqref{CCM summand}. This is consistent with the fact that the $3jm$ symbols have been introduced in \eqref{CCM summand} in order to obtain invariance under moves in the boundary.

In the derivation of the logarithmic corrections to the BTZ black hole entropy via the Euclidean (continuum) path integral \cite{GKS}, a crucial role is played by the requirement of modular invariance at the boundary of the solid torus. It has been suggested in \cite{Sphd} that a realization of this invariance property at the level of the discretized partition function \textit{\`a la} Turaev--Viro could be encoded in the requirement of invariance under boundary moves. This can be seen as an argument motivating the model of \cite{CCM} that we have just described, but further work is required in order to establish a precise relationship between the role of the symmetries in the continuum and at the discrete level.

\subsection{State sum observables}
\label{subsec:observables}

\noindent Finally, we conclude this section on the Turaev--Viro model by introducing, following \cite{BGIM}, the notion of observable. This is the definition that we are going to use in order to compute in the next section the black hole horizon observable.
\begin{definition}[State sum observable]\label{observable}
Let $M$ be a compact manifold without boundary, $\Delta$ one of its triangulations, and $\phi:\{e\in\Delta\}\rightarrow\{0,1/2,1,\dots,k/2\}$ an admissible coloring of the edges of the triangulation by half-integer spins. Given a subset $\Gamma$ of $n$ edges of the triangulation $\Delta$, and a coloring $\phi_\Gamma:\{e\in\Gamma\}\rightarrow\{0,1/2,1,\dots,k/2\}$ of these edges by spins, the observable corresponding to $\Gamma$ is defined as
\be\label{observable formula}
\mathcal{Z}(M,\Gamma(j_1,\dots,j_n))=\sum_{\phi|_\Gamma}\mathcal{Z}(M,\Delta),
\ee
where the sum over $\phi|_\Gamma$ means that the spins coloring the edges of $\Gamma$ are held fixed.
\end{definition}
Such an observable is obviously triangulation-dependent, and it depends on a choice of coloring $\phi_\Gamma$ of the edges of $\Gamma$ by spins. This definition also applies to the case in which the graph $\Gamma$ is composed of a union of disjoint edges. Evidently, one has the property that
\be\label{observable sum complete}
\sum_{\phi_\Gamma}\mathcal{Z}(M,\Gamma(j_1,\dots,j_n))=\mathcal{Z}(M),
\ee
that is, if we sum over the colorings of the edges of $\Gamma$ that have been held fixed in $\phi|_\Gamma$, we recover the topological invariant $\mathcal{Z}(M)$. This comes from the fact that, schematically, $\phi=\phi|_\Gamma+\phi_\Gamma$, which means that the coloring of the edges of $\Delta$ is equal to the coloring of the edges of $\Delta\backslash\Gamma$ plus the coloring of the edges of $\Gamma$. Another important property is that the observable partition function $\mathcal{Z}(M,\Gamma(j_1,\dots,j_n))$ does not depend on the the triangulation of $M$ away from $\Gamma$. For the present work, we will need to consider such a graph observable in the case of a manifold with boundary (the solid torus). Since we will consider (as we shall see shortly) a graph $\Gamma$ with no edges lying in the boundary triangulation $\partial\Delta$, the definition \ref{observable} can be straightforwardly extended.

Before finishing this section, let us briefly recall how the notion of state sum observables given by definition \ref{observable} is related to the Witten-Reshetikhin-Turaev observables. These latter are given (at least formally, since one has to give a meaning to the path integral measure) by the expectation values of knot observables in $\SU(2)$ Chern--Simons theory. For a non-singular knot, the associated Chern--Simons observable is simply given by the trace in a finite-dimensional $\SU(2)$ representation of the holonomy of the Chern--Simons connection along the path defining the knot. When the knot is singular (and therefore has vertices), the associated observable can be constructed in the same way as spin networks are constructed, i.e. by assigning representations to the edges and intertwiners to the vertices. However, contrary to what happens in $\SU(2)$ LQG, in Chern--Simons theory the braiding is relevant and has to be taken into account. Witten was the first to give a precise meaning to the expectation values of these observables \cite{Witten}, and then Reshetikhin and Turaev showed how to relate Witten's construction to the representation theory of $\text{U}_q(\su(2))$ \cite{RT1,RT2}. The calculation \textit{\`a la} Witten-Reshetikhin-Turaev of a colored (at most trivalent) knot observable $\Gamma$ leads to a knot invariant, denoted by $\mathcal{Z}_\text{WRT}(M,\Gamma(\theta_1,\dots,\theta_n))$ in \cite{BGIM}. The representations $(\theta_1,\dots,\theta_n)$ color the $n$ edges of the knot $\Gamma$, and there is no color assigned to the vertices if these are at most trivalent (this can be straightforwardly generalized to any $v$-valent knot \cite{BGIM}).

The relation between the Turaev--Viro invariant of a colored (at most trivalent) knot $\Gamma(j_1,\dots,j_n)$ as defined above (in definition \ref{observable})  and $\mathcal{Z}_\text{WRT}(M,\Gamma(\theta_1,\dots,\theta_n))$ is given by \cite{BGIM}
\ba\label{FT TV to WRT}
\sum_{j_1,\dots,j_n}\mathcal{Z}(M,\Gamma(j_1,\dots,j_n))K_{j_1}(\theta_1)\dots K_{j_n}(\theta_n)&=&
\mathcal{Z}_\text{R}(M,\Gamma(\theta_1,\dots,\theta_n))\nonumber\\
&=&\f{\big|\mathcal{Z}_\text{WRT}(M,\Gamma(\theta_1,\dots,\theta_n))\big|^2}{\displaystyle\prod_v\Theta_v},
\ea
where $\mathcal{Z}_\text{R}(M,\Gamma(\theta_1,\dots,\theta_n))$ is the so-called relativistic spin network invariant, and $\Theta_v$ is related to the normalization of the intertwiner at the vertices $v\in\Gamma$ (we will compute explicitly this normalization factor in the case of the BTZ observable partition function). The kernel $K_j(\theta)$ of this Fourier transform is given by
\be
(-1)^{2\theta}K_j(\theta)=\f{S_j(\theta)}{\omega_j^2}=\f{\displaystyle\sin\left(\f{\pi}{k+2}d_jd_\theta\right)}{\displaystyle\sin\left(\f{\pi}{k+2}d_j\right)},
\ee
where $d_j=2j+1$ is the dimension of the spin $j$ representation, and $S_j(\theta)=(-1)^{2(j+\theta)}\big[d_jd_\theta\big]_q$ are the Verlinde coefficients defined by the evaluation of the Hopf link embedded in $\mathbb{S}_3$. It is immediate to see that the kernel satisfies the symmetry
\be
\sin\left(\f{\pi}{k+2}d_j\right)K_j(\theta)=\sin\left(\f{\pi}{k+2}d_\theta\right)K_\theta(j).
\ee
The relation \eqref{FT TV to WRT} generalizes \eqref{WRT} and establishes that the Turaev--Viro invariant is somehow the Fourier transform of the (modulus squared of the) Witten-Reshetikhin-Turaev invariant. By virtue of the orthogonality
\be
\sum_{j=0}^{k/2}S_j(\theta)S_j(\theta')=\omega^2\delta_{\theta,\theta'}
\ee
of the Verlinde coefficients, the Fourier kernel $K_j(\theta)$ satisfies the orthogonality relation
\be
\sum_{j=0}^{k/2}\sin^2\left(\f{\pi}{k+2}d_j\right)K_j(\theta)K_j(\theta')=\f{k+2}{2}\delta_{\theta,\theta'}.
\ee
This relation allows to compute the inverse Fourier transform and to establish the inverse relation to \eqref{FT TV to WRT}, where $\mathcal{Z}(M,\Gamma(j_1,\dots,j_n))$ is expressed in term of $\mathcal{Z}_\text{WRT}(M,\Gamma(\theta_1,\dots,\theta_n))$. This inverse formula is given explicitly by
\be\label{FT WRT to TV}
\mathcal{Z}(M,\Gamma(j_1,\dots,j_n))=
\omega^{-2n}\left(\prod_{i=1}^n\omega_{j_i}^2\right)\sum_{\theta_1,\dots,\theta_n}\mathcal{Z}_\text{R}(M,\Gamma(\theta_1,\dots,\theta_n))\prod_{i=1}^nS_{j_i}(\theta_i).
\ee
In \cite{FNR}, these results were used to establish interesting duality relations between the $6j$ symbols.

\section{The Euclidean BTZ black hole in the Turaev--Viro model}
\label{sec4}

\noindent In this section we construct the spin foam description of the Euclidean BTZ black hole. For this, we start with the Turaev--Viro model on a solid torus, and introduce an observable (in the sense of \ref{observable}) corresponding to a graph $\Gamma_n$ that tessellates the $\mathbb{S}_1$ circle representing the horizon at the core of the solid torus. We choose our triangulation such that the horizon is tessellated by a succession of $n$ edges colored with representations $j_1,\dots,j_n$ of the quantum group $\text{U}_q(\su(2))$. As a consequence, the Turaev--Viro observable associated to the horizon is a function $\mathcal{Z}(\Gamma_n(j_1,\dots,j_n))$ of these $n$ spins. Using elementary Pachner moves, we then show that $\mathcal{Z}(\Gamma_n(j_1,\dots,j_n))$ is related by a recursion relation to $\mathcal{Z}(\Gamma_{n-1}(j_1,\dots,j_{n-1}))$. This recursion relation can be solved explicitly, and the solution turns out to be, up to a normalization factor, the dimension of the space of $\text{U}_q(\su(2))$-invariant tensors in the tensor product $j_1\otimes\dots\otimes j_n$. This matches the result obtained in \cite{FGNP-BTZ} in the context of canonical quantization. The analytic continuation to $\Lambda<0$ can then be carried out, and the large spin behavior of the (normalized and analytically-continued) Turaev--Viro observable partition function leads to the Bekenstein--Hawking entropy.

As seen in section \ref{sec2}, the Euclidean BTZ black hole spacetime is a solid torus whose core is the horizon. Since the solid torus is a manifold with a boundary, the expression \eqref{TV summand with boundary} is a good starting point to construct a spin foam model description of the BTZ black hole. However, it is not enough to simply write the Turaev--Viro model on a solid torus (which would simply lead to the corresponding invariant), because one has to somehow specify the presence of a horizon in order to obtain a description of a black hole. This can be done by using the notion of state sum observable introduced above. Evidently, this notion breaks the topological invariance of the model, since it is based on the choice of a fixed graph (here representing a tessellation of the horizon). This is however analogous to what happens in the continuum path integral, where one must not only fix the spacetime topology to be that of a solid torus, but also specify kinematical data corresponding to a black hole solution, namely the mass $M$ and in the rotating case also the angular momentum $J$. The ``area'' of the horizon, i.e. the length $L=2\pi r_+$ of the circle at the core of the torus, where the Euclidean radius $r_+$ is given in \eqref{euclidean radius}, must be fixed. This is clearly a geometrical data which naturally breaks the topological invariance of the partition function. The consequence of this simple fact is important, since different choices of triangulation $\Delta$ for the manifold $M$ and of graph $\Gamma$ discretizing the horizon will generically lead to different partition functions. As we shall see, these will however possess the same semiclassical behavior.

\begin{center}
\begin{figure}[h]
\includegraphics[scale=0.5]{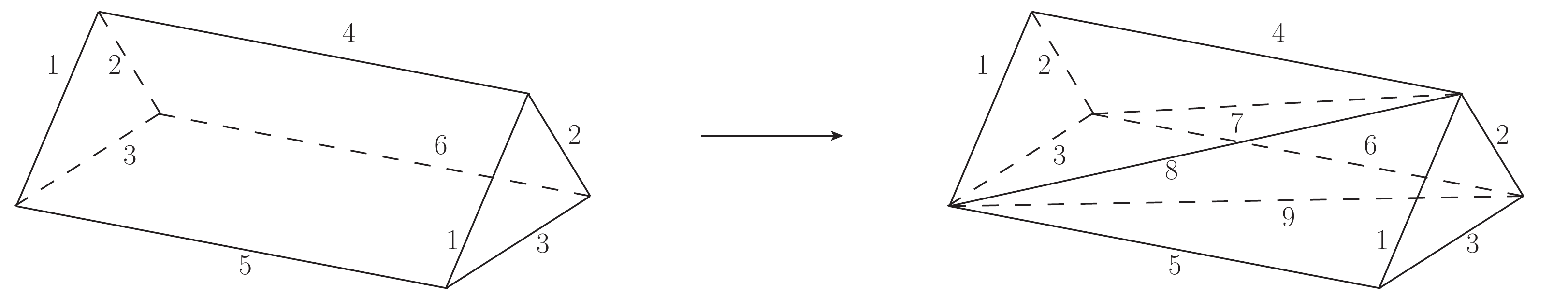}
\caption{Decomposition of a prism into three tetrahedra. The two opposite triangles labelled $(1,2,3)$ have been identified in order to obtain a triangulation of the solid torus.}
\label{triang1}
\end{figure}
\end{center}

The first step is therefore to choose an appropriate triangulation of the solid torus. The simplest triangulation can be obtained by taking a prism with identified opposite triangular faces, and dividing it into three tetrahedra. This triangulation of the solid torus is represented on figure \ref{triang1}.
One obvious drawback of this triangulation is that it has no internal edges. Therefore, although it is a perfectly valid triangulation if one is interested in computing the Turaev--Viro invariant of the solid torus, it is not quite appropriate if one wants to use it to define an observable associated to the horizon. Indeed, the horizon being a circle at the core of the solid torus, it is natural to choose a triangulation with internal edges, and then to define a graph triangulating the horizon in term of a subset of these internal edges. For this reason, we are going to consider a triangulation of the solid torus which naturally induces a discretization of the horizon into $n$ internal edges. Then, if one colors the edges of this triangulation by spins, the status of the spins coloring the edges of the discretized horizon is radically different from that of the spins coloring all the other edges. Indeed, the definition \ref{observable} of the observable partition function implies that the first set of spins is fixed once and for all, while the second one is summed over. The fixed spins $j_e$ labeling the edges of the discretized horizon are then associated with a quantum of length $L_e=8\pi\lp\sqrt{j_e(j_e+1)}$, and the sum of these contributions gives the macroscopic length of the horizon. Let us now carry out this construction explicitly.
 
\subsection{Choice of triangulation}

\begin{center}
\begin{figure}[h]
\includegraphics[scale=0.5]{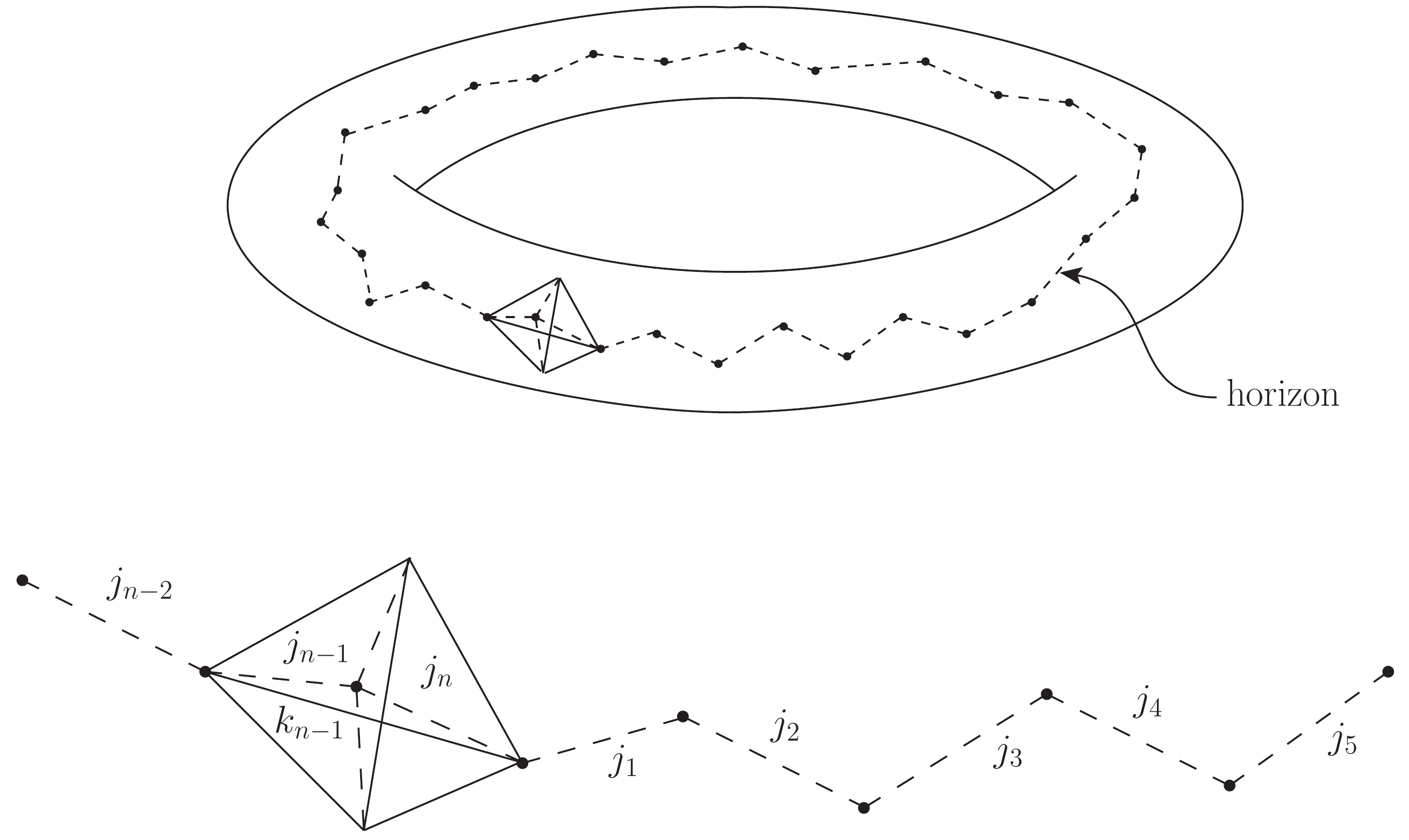}
\caption{Top: Solid torus with a portion of triangulation of its interior, and the dashed line representing the discretized horizon graph $\Gamma_n$. Bottom: Detail of the tessellation of the horizon into $n$ edges labelled by spins $j_1,\dots,j_n$. We have represented a tetrahedron surrounding the vertex between the edges $j_{n-1}$ and $j_n$ since later on we will use a partial Pachner 4--1 move to replace $j_{n-1}$ and $j_n$ by $k_{n-1}$.}
\label{local discretization}
\end{figure}
\end{center}

\noindent We start by discussing further our choice of triangulation of the solid torus, and the subsequent description of the tessellated horizon. The key point is to realize that the choice of triangulation away from the horizon is irrelevant for the calculation of the entropy. Therefore, one can start by considering an even number $n$ (with $n>2$) of edges linked together and forming a closed one-dimensional graph $\Gamma_n$ running all around the interior of the solid torus. These edges form a discretization of the horizon. Then one simply has to choose a triangulation of the solid torus compatible with this discretization of the horizon. The only requirement that we impose is that the edges forming the horizon never meet the boundary triangulation. One can therefore choose any triangulation which looks like the one represented on figure \ref{local discretization}, i.e. which is sufficiently refined for no vertices of the horizon to touch the boundary. 

With the choice of triangulation $\Delta$ represented on figure \ref{local discretization} and the definitions \eqref{TV summand with boundary} and \eqref{observable formula}, we can now define the observable partition function associated to the horizon. It is given by
\bas
\mathcal{Z}(\Gamma_n(j_1,\dots,j_n))&=&\sum_{\phi|_{\Gamma_n}}\prod_{v_b}\omega^{-1}\prod_{v_i}\omega^{-2}\prod_{e_b}\omega_{j_e}\prod_{e_i}\omega_{j_e}^2\prod_\tau\big|6j_{e\subset\tau}\big|_q\label{BTZ partial}\\
&=&\sum_{\phi|_{\Gamma_n}}\prod_{v_b}\omega^{-1}\prod_{v_i\backslash v_{\Gamma_n}}\omega^{-2}\prod_{v_{\Gamma_n}}\omega^{-2}\prod_{e_b}\omega_{j_e}\prod_{e_i\backslash v_{\Gamma_n}}\omega_{j_e}^2\prod_{e_{\Gamma_n}}\omega_{j_e}^2\prod_\tau\big|6j_{e\subset\tau}\big|_q,\qquad\label{BTZstatesum}
\eas
where the sum is taken over the colorings $\phi|_{\Gamma_n}$ that do not affect the horizon edges $e_{\Gamma_n}$. In the second line we have simply made the explicit distinction between the internal edges and vertices that belong to the horizon, and the ones that do not belong to the horizon.

\subsection{Calculation of the partition function}

\noindent We are now going to show that our choice of triangulation allows for the partition function \eqref{BTZstatesum} to be computed explicitly. This computation relies on the fact that $\mathcal{Z}(\Gamma_n(j_1,\dots,j_n))$ satisfies a certain recursion relation. To see that this is indeed the case, one has to use the following property satisfied by the quantum $6j$ symbols.
\begin{property}[Partial 4--1 Pachner move] Using the graphical correspondence between tetrahedra and quantum $6j$ symbols, we have
\be\label{lemma}
\sum_{k,l}\omega_k^2\omega_l^2\omega_{j_1}^2\omega_{j_2}^2\vcenter{\hbox{\includegraphics[scale=0.5]{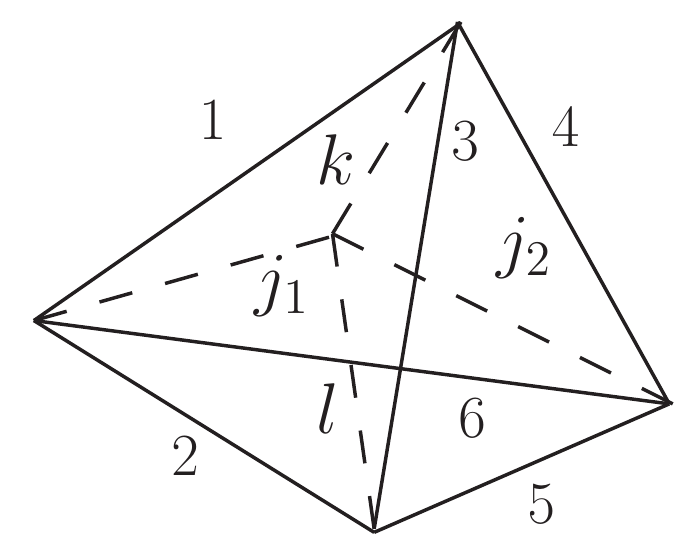}}}
=
\omega_{j_1}^2\omega_{j_2}^2\omega^{-2}_6Y(j_1,j_2,6)\vcenter{\hbox{\includegraphics[scale=0.5]{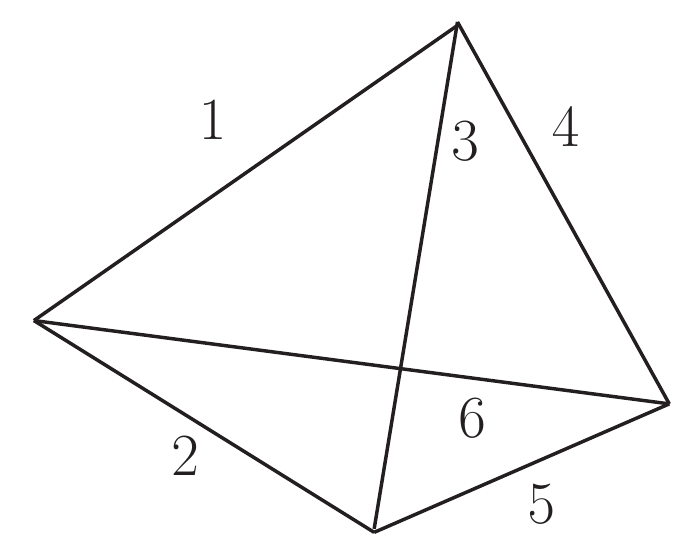}}}.
\ee
\end{property}
\begin{proof}
The formula for the partial 4--1 Pachner move is a simple consequence of the Biedenharn-Elliot identity \eqref{BE} and the orthogonality relation \eqref{ortho}. By writing explicitly the quantum $6j$ symbols associated to the four tetrahedra on the left-hand side, one gets
\begin{subequations}
\ba
&&\sum_{k,l}\omega_k^2\omega_l^2\omega_{j_1}^2\omega_{j_2}^2
\left|
\begin{array}{ccc}
1&3&2\\
l&j_1&k
\end{array}\right|_q
\left|
\begin{array}{ccc}
3&4&5\\
j_2&l&k
\end{array}\right|_q
\left|
\begin{array}{ccc}
2&5&6\\
j_2&j_1&l
\end{array}\right|_q
\left|
\begin{array}{ccc}
4&1&6\\
j_1&j_2&k
\end{array}\right|_q\label{sum1}\\
&=&\sum_k\omega_k^2\omega_{j_1}^2\omega_{j_2}^2
\left|
\begin{array}{ccc}
5&2&6\\
1&4&3
\end{array}\right|_q
\left|
\begin{array}{ccc}
6&j_1&j_2\\
k&4&1
\end{array}\right|_q
\left|
\begin{array}{ccc}
4&1&6\\
j_1&j_2&k
\end{array}\right|_q\\
&=&\omega_{j_1}^2\omega_{j_2}^2\omega^{-2}_6Y(j_1,j_2,6)
\left|
\begin{array}{ccc}
5&2&6\\
1&4&3
\end{array}\right|_q,\label{sum3}
\ea
\end{subequations}
which establishes the proof.
\end{proof}
To emphasize the consistency of this relation which can seem surprising at first (since it shows that only half of the internal summation is needed in order to collapse the four tetrahedra into one), one can use the formula
\be\label{condition weights}
\omega^2=\omega_j^{-2}\sum_{k,l}\omega_k^2\omega_l^2 Y(j,k,l),
\ee
where the sum is taken over the values of $k$ and $l$ which are such that the triple $(j,k,l)$ is admissible. This simply means that if we were to sum \eqref{sum1} over $j_1$ and $j_2$ as well, we would complete the partial 4--1 Pachner move and end up with the tetrahedron on the right of \eqref{lemma} with the weight $\omega^2$. This weight would then kill one of the weights $\omega^{-2}$ attached to an internal vertex in the partition function \eqref{BTZ partial}. In the absence of the horizon observable, this fact is responsible for the invariance of the partition function under the 4--1 Pachner moves. As mentioned above, we see that the introduction of an observable in the partition function breaks the topological invariance, since fixing the spins labeling the horizon enables us only to perform partial Pachner moves in the vicinity of the horizon.

We can now start to perform a succession of partial Pachner moves in the observable partition function \eqref{BTZ partial}. After a first partial Pachner move in the tetrahedron surrounding the edges $(j_{n-1},j_n,k_{n-1})$ of figure \ref{local discretization}, we obtain
\be\label{recursion relation}
\mathcal{Z}(\Gamma_n(j_1,\dots,j_n))=\omega^{-2}\sum_{k_{n-1}}\omega_{j_{n-1}}^2\omega_{j_n}^2\omega_{k_{n-1}}^{-2}Y(j_{n-1},j_n,k_{n-1})
\mathcal{Z}(\Gamma_{n-1}(j_1,\dots,j_{n-2},k_{n-1})).
\ee
This recursion relation relates the partition function based on a triangulation of the horizon into $n$ edges with the one based on a triangulation of the horizon into $(n-1)$ edges. Using this formula recursively, one arrives immediately at the following expression for the partition function:
\bas\label{TVBTZintermediate}
\mathcal{Z}(\Gamma_n(j_1,\dots,j_n))&=&\omega^{-2(n-1)}\sum_{k_1,\dots,k_{n-1}}\left(\prod_{i=1}^n\omega_{j_i}^2\right)Y(j_{n-1},j_n,k_{n-1})\left(\prod_{i=1}^{n-2}Y(j_i,k_{i+1},k_i)\right)\omega_{k_1}^{-2}\mathcal{Z}(\Gamma_1(k_1))\nonumber\\
&=&\omega^{-2(n-1)}\sum_{k_1,\dots,k_{n-1}}\left(\prod_{i=1}^n\omega_{j_i}^2\right)\left(\prod_{i=1}^{n-1}Y(j_i,k_{i+1},k_i)\right)\delta_{j_n,k_n}\omega_{k_1}^{-2}\mathcal{Z}(\Gamma_1(k_1)).
\eas
We obtain the result that the partition function is completely determined by $\mathcal{Z}(\Gamma_1(k_1))$, which is the observable partition function for a graph with only one edge and one vertex.

One should now proceed with the computation of $\mathcal{Z}(\Gamma_1(k_1))$. To do so, the triangulation of the solid torus has to be chosen in such a way that there is an internal edge (which does not meet the boundary) going around the torus and closing onto itself. The simplest triangulation which has only one internal edge $\Gamma_1$ that does not meet the boundary can be obtained by gluing together three of the prisms represented on the right of figure \ref{triang1}. Then, in order to evaluate $\mathcal{Z}(\Gamma_1(k_1))$ as a function of $k_1$ on this triangulation, one needs to make a choice between the two possible versions \eqref{TV summand with boundary} or \eqref{CCM summand} of the partition function for a manifold with boundaries. In this sense, the evaluation of $\mathcal{Z}(\Gamma_1(k_1))$ depends on how we treat the boundary of the solid torus and on which conditions are imposed thereon. Different choices for the Turaev--Viro partition function on a manifold with boundary will naturally impact the evaluation of $\mathcal{Z}(\Gamma_1(k_1))$.

To summarize the situation, we have only considered up to now what happens in the vicinity of the horizon in order to establish the recursion relation \eqref{recursion relation} and to arrive at the expression \eqref{TVBTZintermediate}, and it appears that to complete the calculation and find the final expression for the partition function we now need to specify particular conditions on the boundary. However, we are going to see that the explicit expression for $\mathcal{Z}(\Gamma_1(k_1))$ is not necessary in order to obtain the semiclassical behavior of the partition function and to recover the Bekenstein--Hawking entropy formula (once the analytic continuation to $\Lambda<0$ is performed). 

In order to better understand the role of $\mathcal{Z}(\Gamma_1(k_1))$, let us go back to the computation made in \cite{FGNP-BTZ} in the context of canonical quantization. In the canonical framework described in \cite{FGNP-BTZ}, the spin network states at fixed instant of time span the entire spatial slice, and are not define only in the vicinity of the horizon as it is the case in four-dimensional LQG. This has in fact important consequences, since it shows that the space-like $\mathbb{S}_1$ boundary at spatial infinity (recall that the constant time hypersurfaces correspond to $\theta=\text{constant}$ slices on figure \ref{BTZrepresentation}) is somehow related to the horizon, which has an $\mathbb{S}_1$ topology as well. Indeed, one can choose the canonical states to be supported on graphs with edges crossing the horizon as well as the $\mathbb{S}_1$ boundary at the infinity. The consequence of such a choice of quantum canonical states implies particular boundary conditions at spatial infinity. The knowledge of these conditions is not needed in order to proceed with our analysis, but it is however clear that they should relate the mass of the black hole as measured at the horizon (through the measure of the length for instance) and the mass measured at the infinity (the ADM mass for instance). In order for the results of canonical and spin foam quantization to agree, we have to define $\mathcal{Z}(\Gamma_1(k_1))$ in such a way that the Turaev--Viro observable reproduce the number of states $\mathcal{N}(j_1,\dots,j_n)$ that was derived in \cite{FGNP-BTZ} in the context of canonical quantization (up to an eventual normalization factor). The expression for this number of states $\mathcal{N}(j_1,\dots,j_n)$  is recalled in the following section in equation \eqref{dimension}. It is however very useful at this point to write $\mathcal{N}(j_1,\dots,j_n)$ as
\ba
\mathcal{N}(j_1,\dots,j_n)&=&\sum_{k_1,\dots,k_{n-1}}\delta_{j_n,k_n}\delta_{k_1,0}\prod_{i=1}^{n-1}Y(j_i,k_{i+1},k_i)\nonumber \\
&=&\sum_{k_1,\dots,k_{n-1}}\delta_{j_n,k_n}\delta_{k_1,0}\delta_{k_{n+1},0}Y(j_n,k_{n+1},j_n)\prod_{i=1}^{n-1}Y(j_i,k_{i+1},k_i)\nonumber\\
&=&\sum_{k_1,\dots,k_n}\delta_{k_1,0}\delta_{k_{n+1},0}\prod_{i=1}^nY(j_i,k_{i+1},k_i)\label{dim formula},
\ea
where the last equality is the usual formula for the dimension of the Chern--Simons Hilbert space \cite{KMdim,ENPP}. This rewriting is possible since for every $j_n\leq k/2$ (where $k$ is the Chern--Simons level) the triple $(j_n,0,j_n)$ is admissible in the sense of \eqref{admin} and we have $Y(j_n,0,j_n)=1$. As a consequence, in order for the expressions \eqref{dim formula} and \eqref{TVBTZintermediate} to agree, $\mathcal{Z}(\Gamma_1(k_1))$ has to be vanishing if the representation $k_1$ is non-trivial. We can therefore write that $\mathcal{Z}(\Gamma_1(k_1))=\mathcal{Z}(\mathbf{T})\delta_{k_1,0}$ where $\mathcal{Z}(\mathbf{T}) $ is the Turaev--Viro invariant on the solid torus without any representation observable inserted. As a conclusion, we see that there is a particular choice of $\mathcal{Z}(\Gamma_1(k_1))$ (i.e. a particular treatment of the boundary data) which leads to an observable partition function compatible with the derivation of the number of states of \cite{FGNP-BTZ}, and for which we finally obtain
\be\label{number of states SF}
\mathcal{Z}(\Gamma_n(\vec{\jmath}_n))=\omega^{-2(n-1)}\mathcal{Z}(\mathbf{T})\mathcal{N}(\vec{\jmath}_n)\prod_{i=1}^n\omega_{j_i}^2,
\ee
where $\vec{\jmath}_n$ is a shorthand notation for the collection of $n$ spins.

Before going on, let us discuss a bit further the significance of $\mathcal{Z}(\Gamma_1(k_1))$. From the formula \eqref{FT TV to WRT}, one can see that $\mathcal{Z}(\Gamma_1(k_1))$ is related to the expectation value $\la\mathcal{O}(\theta)\ra$ \textit{\`a la} Witten-Reshetikhin-Turaev of the Wilson loop $\mathcal{O}$ along the horizon colored by a finite-dimensional representation $\theta$. This relation is given by
\be
\sum_{j=0}^{k/2}\mathcal{Z}(\Gamma_1(j))K_j(\theta)=\f{\la\mathcal{O}(\theta)\ra^2}{\omega_\theta^2},
\ee
where we have used the normalization condition \eqref{loop normalization}, and removed the absolute value because the expectation value $\la\mathcal{O}(\theta) \ra$ is real. As a consequence of this equation, we see that the condition that $\mathcal{Z}(\Gamma_1(j))$ vanishes if the representation $j$ is non-trivial implies naturally that $\la\mathcal{O}(\theta) \ra$ is proportional to the quantum dimension $\big[d_\theta\big]_q$ (as a function of $\theta$).  

In order to further emphasize this fact, let us propose a new proof of \eqref{number of states SF} starting from the Fourier transform relation \eqref{FT WRT to TV}. When applied to the graph $\Gamma_n$ which discretizes the black hole horizon, one sees immediately that the partition function is defined by
\ba\label{relativistic graph}
\omega^{2n}\left(\prod_{i=1}^n\omega_{j_i}^{-2}\right)\mathcal{Z}(\Gamma_n(\vec{\jmath}_n))
&=&\sum_{\theta_1,\dots,\theta_n}\mathcal{Z}_\text{R}(M,\Gamma(\theta_1,\dots,\theta_n))\prod_{i=1}^nS_{j_i}(\theta_i)\nonumber\\
&=&\sum_{\theta=0}^{k/2}\f{\la\mathcal{O}(\theta)\ra^2}{\omega_\theta^{2n}}\prod_{i=1}^nS_{j_i}(\theta),
\ea
where we have successively used the relation
\be\label{loop normalization}
\Big\langle\vcenter{\hbox{\includegraphics[scale=0.5]{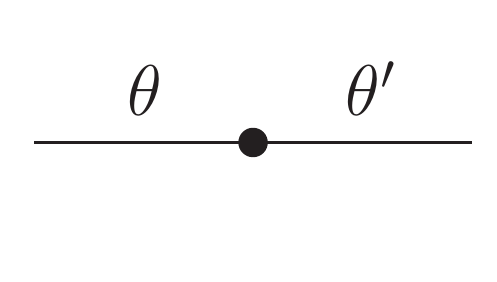}}}\Big\rangle_\text{R}
=\f{1}{\omega_\theta^2}\delta_{\theta,\theta'}
\Big\langle\vcenter{\hbox{\includegraphics[scale=0.5]{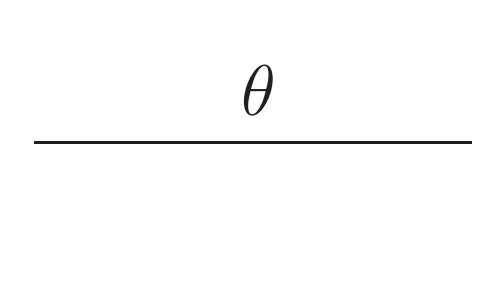}}}\Big\rangle_\text{R}
\ee
between the relativistic spin network evaluations to replace the graph $\Gamma_n$ with $n$ edges by a Wilson loop $\Gamma_1=\mathcal{O}$. We now see that when the expectation value is
\be
\la\mathcal{O}(\theta)\ra^2=\f{2}{k+2}\omega^2\sin^2\left(\f{\pi}{k+2}d_\theta\right),
\ee
the previous expression agrees with the expression found in the canonical formulation, i.e. formula \eqref{dimension} (or equivalently \eqref{dim formula}). This is consistent with the observation that for the spin foam and canonical quantizations to agree, $\la\mathcal{O}(\theta)\ra$ must be proportional to the quantum dimension $\big[d_\theta\big]_q$ as a function of $\theta$. Furthermore, we can see on \eqref{relativistic graph} that the choice of boundary conditions affects only the measure factor in the sum over $\theta$, and not the product of quantum characters (or Verlinde coefficients). It is however this latter that is responsible for the leading order semiclassical behavior of the partition function. We can therefore conclude that the precise evaluation of $\mathcal{Z}(\Gamma_1(j))$ is not essential for the derivation of the entropy.

Let us finish this section with a remark. When defining the state sum observable \ref{observable} and then applying this definition to describe the horizon, we have assumed that the edges and vertices of the graph $\Gamma$ have the same weights as the corresponding simplices in the bulk of $\Delta\backslash\Gamma$, namely $\omega^2_j$ for the edges and $\omega^{-2}$ for the vertices. However, the status of the weights assigned to the graph $\Gamma$ is quite unclear, since topological invariance cannot be used to determine what they should be once the graph state sum observable is considered. In fact, the only condition that can motivate the use of the original weights $\omega^2_j$ and $\omega^{-2}$ is \eqref{observable sum complete}, which indeed requires for the weights to be left unchanged. If one chooses to drop this requirement, the weights assigned to $\Gamma$ can a priori be chosen arbitrarily. As we will discuss in subsection \ref{sec:log}, the precise form of these weights does not affect the leading order contribution to the entropy, but might be important for the derivation of the subleading (and possibly logarithmic) corrections.

\subsection{Entropy}

\noindent The proof that $\mathcal{N}(\vec{\jmath})$ reproduces the Bekenstein--Hawking entropy in the case $\Lambda<0$ is essentially the same as in \cite{FGNP-BTZ}. One starts with the formula for the dimension of the invariant Hilbert space that was derived in \cite{KMdim}:
\ba
\mathcal{N}(\vec{\jmath})&=&
\f{2}{k+2}\sin^2\left(\f{\pi}{k+2}\right)\sum_{\theta=0}^{k/2}\big[d_\theta\big]_q^{2-n}\prod_{e=1}^nS_{j_e}(\theta)\nonumber\\
&=&\f{2}{k+2}\sum_{d=1}^{k+1}\sin^{2-n}\left(\f{\pi}{k+2}d\right)
\prod_{e=1}^n\sin\left(\f{\pi}{k+2}dd_{j_e}\right),\label{dimension}
\ea
where again $d_{j_e}=2j_e+1$ is the classical dimension of the spin $j_e$ representation coloring the edge $e$ of the horizon, and $S_{j_e}(\theta)$ are the Verlinde coefficients. When the level $k$ is large, one can make the approximation $k+2\sim k+1\sim k$, and then perform the analytic continuation to a negative value of $\Lambda$ by setting $k=i\lambda$ with $\lambda >0$ (see \eqref{q and k}). In order for this analytic continuation to make sense, one must use $|k|=\lambda$ as the upper bound in the sum \eqref{dimension} and in the prefactor. One then obtains
\be\label{N BTZ}
\mathcal{N}_\text{BTZ}(\vec{\jmath})\simeq\f{2}{\lambda}\sum_{d=1}^\lambda\sinh^{2-n}\left(\f{\pi}{\lambda}d\right)
\prod_{e=1}^n\sinh\left(\f{\pi}{\lambda}dd_{j_e}\right),
\ee
where the subscript BTZ has been added to denote that this quantity can be seen as a number of states in the Turaev--Viro model with a negative cosmological constant.
In the semiclassical limit where the spins $j_e$ become large and $\lp$ approaches zero with the product $\lp j_e$ remaining finite, the sum \eqref{N BTZ} is dominated by the term $d=\lambda$, and one gets, using the length relation
\be\label{length relation}
L=8\pi\lp\sum_{e=1}^n\sqrt{j_e(j_e+1)},
\ee
a leading order entropy contribution of
\be
S=\log\mathcal{N}_\text{BTZ}(\vec{\jmath})=\f{L}{4\lp}+\circ\big(\log L\big).
\ee
Here, the logarithmic correction to the semiclassical limit is due to the factor of $\lambda$ in the denominator of the quantity \eqref{N BTZ}. However, as we will discuss it in section \ref{sec:log}, we do not know the precise form of the logarithmic corrections to our analytically-continued model.

\section{Assorted comments}
\label{sec5}

\noindent In this section, we discuss the relationship between our result and the derivation of black hole entropy in four-dimensional LQG, and briefly comment on the subleading corrections.

\subsection{The Barbero--Immirzi parameter and four-dimensional black holes in LQG}
\label{subsec:4d}

\noindent It is interesting to compare our derivation of the Bekenstein--Hawking relation with that of four-dimensional LQG \cite{BHentropy1,BHentropy2,BHentropy3,BHentropy4,BHentropy5,BHentropy6}. The description of black holes in four-dimensional LQG relies on the fact that the symplectic structure of first order gravity, when written in terms of the Ashtekar--Barbero connection, induces an isolated horizon $\SU(2)$ Chern--Simons theory. The entropy in the microcanonical ensemble, where the macroscopic area $A_\text{H}$ of the horizon is held fixed, is then obtained as the logarithm of the number of microstates $\mathbf{N}_\gamma(A_\text{H})$, which is related to the dimension \eqref{dimension} of the Chern--Simons Hilbert space by
\be\label{4d states}
\mathbf{N}_\gamma(A_\text{H})=\sum_{p=0}^\infty\sum_{j_1,\dots,j_p}\delta\left(A_\text{H}-8\pi\gamma\lp^2\sum_{i=1}^p\sqrt{j_i(j_i+1)}\right)\mathcal{N}(j_1,\dots,j_p),
\ee
where $\lp=\sqrt{G\hbar}$ is now the four-dimensional Planck length, and $\gamma$ is the Barbero--Immirzi parameter. The meaning of this formula is that one has to sum the dimension $\mathcal{N}(j_1,\dots,j_p)$ over all the configurations compatible with the macroscopic area $A_\text{H}$, i.e. over all the possible numbers $p$ of (distinguishable) punctures and all the values of the spins. One can then obtain the Bekenstein--Hawking relation by fixing the Barbero--Immirzi parameter to a particular value $\gamma_0\in\mathbb{R}$, and show that the contributions to the entropy come essentially from small spins.

Clearly, this derivation is different from the one we have just presented for the BTZ black hole (or equivalently the one of \cite{FGNP-BTZ}). Indeed, we have obtained the entropy law from the analytic continuation of the Chern--Simons dimension $\mathcal{N}$ alone, without summing over the number $n$ of edges discretizing the horizon (which is the analogue of $p$), without fixing in the length spectrum a free parameter like $\gamma$ to a particular value, and furthermore by taking the large spin limit.

This observation has to be put in parallel with the result of \cite{FGNP-4d}, which shows that in four-dimensional LQG the analytically-continued quantity \eqref{N BTZ} naturally appears if one works with the self-dual value $\gamma=i$ of the Barbero--Immirzi parameter. However, in this case again the entropy can be derived with a fixed number of punctures, and requires to take the large spin limit. Therefore, it is tempting to say that in four dimensions there is a duality between two alternatives:
\begin{itemize}
\item[i)] Working with \eqref{4d states}, which amounts to summing over the punctures and the spins, requires to fix $\gamma=\gamma_0$, and implies small spin domination.
\item[ii)] Working with the analytic continuation of \eqref{dimension}, which can be done with a fixed number of punctures, requires to consider the large spin limit, and can be interpreted as choosing the self-dual value $\gamma=i$.
\end{itemize}
Although this second alternative suggests the possibility of interpreting the Barbero--Immirzi parameter as a regulator that is introduced in order to construct the kinematical structure of the theory with a compact gauge group, and that can be appropriately removed by going back to the value $\gamma=i$ defining the original complex Ashtekar connection \cite{Ashtekar}, the physical meaning of this procedure is still unclear and has to be investigated further. What is however clear is that for the description of the BTZ black hole that we have presented, and for the canonical calculation of \cite{FGNP-BTZ}, the analytic continuation from \eqref{dimension} to \eqref{N BTZ} is a necessary step since it encodes the passage from $\Lambda>0$ to $\Lambda<0$. This seems to indicate that the quantity \eqref{N BTZ} might encode some universal properties of black hole entropy in both three and four dimensions. Furthermore, in light of the results of \cite{Bodendorfer} which show that the state counting for higher-dimensional black holes reduces to the four-dimensional picture described above in i), one can conjecture that the analytically-continued quantity \eqref{N BTZ} does actually encode information about black hole entropy in any dimension $d\geq3$ \cite{GNP}.

In order to further clarify the role played by the Barbero--Immirzi parameter, it is interesting to look at the result of \cite{BGNY}. There, it was shown that first order three-dimensional Euclidean and Lorentzian gravity can both be written as $\SU(2)$ theories, with an Ashtekar--Barbero connection, and scalar and vector constraints analogous to that of the four-dimensional $\SU(2)$ theory. In this case, just like in four dimensions, the kinematical length operator inherits a dependency on the Barbero--Immirzi parameter, and evidently becomes discrete even in the Lorentzian case. This is already a surprising artifact due to the introduction of the three-dimensional Barbero--Immirzi parameter. Furthermore, although no description of the entropy of a BTZ black hole has been proposed in this framework, it seems quite clear that one should not do so by following the derivation of four-dimensional black hole entropy along the lines of alternative i) described above. Indeed, in the Euclidean case for example (since the physical states of the Lorentzian theory with $\Lambda\neq0$ are not explicitly known), one could think of using the physical states for $\Lambda>0$, which are the $\text{U}_q(\su(2))$ spin network with $q$ a root of unity, and then it is clear in light of the above discussion that one should return to $\Lambda<0$ by analytic continuation, and use the Euclidean self-dual value $\gamma=1$. Alternatively, one could consider fixing $\gamma$ in order to get the correct Bekenstein--Hawking relation, as in \cite{SKG,GI1}, but then one would lose the ability of performing the analytic continuation to $\Lambda<0$.

All these observations support the description of the BTZ black hole entropy that we have proposed in this paper, and bring additional credit to the four-dimensional proposal of \cite{FGNP-4d} and to the idea of working with the self-dual variables \cite{GNP}.

\subsection{Logarithmic corrections}
\label{sec:log}

\noindent One question which we did not address in this work is that of the subleading corrections to the entropy. It is generally accepted that these should be logarithmic with a factor of $-3/2$. This was derived for example from the corrections to the Cardy formula in \cite{Carlip-log}, and from the continuous path integral in \cite{GKS}.

Analyzing the subleading corrections to \eqref{number of states SF} is rather subtle because the corrections to the analytic continuation \eqref{N BTZ} of \eqref{dimension} are not known. The prefactor contributes with a term of the form $-\log\lambda$, which can be written in terms of the horizon length $L$ since we have the relation
\be
k=\f{L}{4\pi\lp\sqrt{GM}},
\ee
but we do not know how the rest of \eqref{N BTZ} scales beyond leading order. Furthermore, it is known that the weights assigned to the vertices can be written as
\be
\omega^2=\f{k^3}{2\pi^2}\big(1+\circ\big(k^{-2}\big)\big),
\ee
but unclear whether these weights along with the $\omega_j$'s should be analytically-continued as well.

As mentioned in section \ref{subsubsec:PL}, it would be interesting to see if the requirement of modular invariance of the boundary of the solid torus imposed in \cite{GKS} can be translated into a condition at the level of the state sum model, and whether this could fix some of the above-mentioned ambiguities by some physical requirements.

\subsection{Relationship with other approaches}

\noindent One very interesting open question is that of the relationship between the present calculation and other proposals for the derivation of the entropy of a BTZ black hole. Most of the knowledge that we have about three-dimensional black holes comes from the techniques of conformal field theory \cite{Carlip}, which our approach seems quite remote from. As argued already in \cite{FGNP-BTZ}, it is natural to see the analytically-continued quantity $\mathcal{N}_\text{BTZ}(\vec{\jmath})$ defined in \eqref{N BTZ} as the analogue of the density of states of conformal field theory. The key to understanding more rigorously this analogy (which for the moment holds only on the basis that these two quantities have the same leading order behavior) would be to study the type of discrete state sum model that is obtained on the boundary of the solid torus. As discussed in section \ref{sec:IIIb}, there are two possible ways of writing the state sum model on the boundary, which correspond to the choices \eqref{TV summand with boundary} and \eqref{CCM summand}, and one should investigate whether the analytic continuation to $\Lambda<0$ can be given a meaning already at the level of these expressions, and whether this procedure can be given a physical interpretation.

Finally, it would be interesting to investigate whether the proposal made in section \ref{subsec:4d} concerning the universality of the analytically-continued Chern--Simons Hilbert space dimension has any relationship with the universality proposed by Carlip and based on conformal field theory \cite{Carlip-CFT1,Carlip-CFT2,Carlip-CFT3,Carlip-CFT4}.

\section{Conclusion and discussion}
\label{sec6}

\noindent In this work, we have derived the Bekenstein--Hawking entropy of a Euclidean BTZ black hole from the Turaev--Viro state sum model. As explained in the introduction, the apparent difficulty in doing so resides in the fact that no spin foam model is known for three-dimensional gravity with a negative cosmological constant, which is a necessary condition for the existence of the BTZ solution. Therefore, we have argued that one possible route to circumvent this problem is to start from the Turaev--Viro model, which represents the spin foam quantization of Euclidean three-dimensional gravity with a positive cosmological constant. By doing so, one can take advantage of the fact that the Euclidean BTZ black hole has the topology of a solid torus, write the Turaev--Viro model on this manifold, and then introduce the notion of a graph observable. This natural notion of observable (which as we have recalled in section \ref{subsec:observables} is related to the usual Chern--Simons observables), when applied to a graph representing a tessellated circle with $n$ edges, can be related by a partial 4--1 Pachner move to the observable defined on a graph with $n-1$ edges. We have shown that the resulting recursion relation leads to the dimension of the Chern--Simons Hilbert space of tensor product between $n$ representations of $\text{U}_q(\su(2))$. This quantity is the same as the one introduced in the canonical framework in \cite{FGNP-BTZ}, and also the key ingredient for the state counting in four-dimensional LQG. In order to go back to the physically relevant situation and be able to talk about an actual black hole, we have proposed an analytic continuation of the Chern--Simons level, which amounts to changing the sign of the cosmological constant from positive to negative. The resulting analytically-continued dimension can therefore be thought of as being associated with the discretized horizon of a BTZ black hole, and upon use of the length relation \eqref{length relation} one can prove that its logarithm reproduces the expected Bekenstein--Hawking relation.

We believe that this result corrects the previous proposals of \cite{SKG,GI1} in two very important ways. First, it shows that the correct factor of $1/4$ in the Bekenstein--Hawking relation can be obtained without having to introduce by hand a Barbero--Immirzi-like parameter in the length spectrum. Second, it explicitly realizes the passage to $\Lambda<0$, which is a necessary condition in order to be able to talk about a BTZ black hole. Furthermore, as discussed in section \ref{subsec:4d}, it is clear that these two facts are related to one another and might have important consequences in four dimensions. In our opinion, it is very interesting to observe that the quantity \eqref{N BTZ} encoding the entropy also appears in four-dimensional LQG once the Barbero--Immirzi parameter is taken to be imaginary.

Finally, we would like to mention that while in this work we did not propose an interpretation for the origin of the microstates contributing to the entropy, our calculation can be seen from the more mathematical side as a way of defining observables in state sum models based on non-compact gauge groups.

\section*{Acknowledgments}

\noindent MG would like to thank Bianca Dittrich for discussions and for sharing a draft of \cite{BK}, Abhay Ashtekar for discussions, and J. Manuel Garc\'ia-Islas and Romesh K. Kaul for correspondence. MG is supported by the NSF Grant PHY-1205388 and the Eberly research funds of The Pennsylvania State University.

\appendix

\section{Properties of the $\boldsymbol{\text{U}_q(\su(2))}$ recoupling symbols}
\label{appendix:6jq}

\noindent In this appendix we recall some useful definitions and properties. First of all, let us define for any integer $n\geq1$ the factorial $[n]_q!=[n]_q[n-1]_q\dots[2]_q[1]_q$, and set $[0]_q!=[0]_q=1$. With this, we can then define for any admissible triple $(i,j,k)$ the quantity
\be
\Delta(i,j,k)=\left(\f{[i+j-k]_q![i+k-j]_q![k+j-i]_q!}{[i+j+k+1]_q!}\right)^{1/2}.
\ee
A triple $(i,j,k)$ is said to be admissible if $Y(i,j,k)=1$, i.e. if the representations $(i,j,k)$ satisfy the conditions \eqref{admin}. 

The Rakah--Wigner quantum $6j$ symbol is then given by the formula \cite{KR}
\ba
\left\{
\begin{array}{ccc}
i&j&k\\
l&m&n
\end{array}\right\}^\text{RW}_q
&=&\Delta(i,j,k)\Delta(i,m,n)\Delta(j,l,n)\Delta(k,l,m)\sum_z(-1)^z[z+1]_q!\nonumber\\
&&\times\f{\Big([i+j+l+m-z]_q![i+k+l+n-z]_q![j+k+m+n-z]_q!\Big)^{-1}}{[z-i-j-k]_q![z-i-m-n]_q![z-j-l-n]_q![z-k-l-m]_q!},\qquad
\ea
where the sum runs over
\be
\max(i+j+k,i+m+n,j+l+n,k+l+m)\leq z\leq\min(i+j+l+m,i+k+l+n,j+k+m+n).
\ee
The quantum $6j$ symbol is defined in terms of the Rakah--Wigner coefficient as
\be
\left|
\begin{array}{ccc}
i&j&k\\
l&m&n
\end{array}\right|_q
\equiv
(\sqrt{-1})^{-2(i+j+k+l+m+n)}
\left\{
\begin{array}{ccc}
i&j&k\\
l&m&n
\end{array}\right\}^\text{RW}_q.
\ee
Since one can associate to a $6j$ symbol a tetrahedron whose edges are colored by the 6 representations involved (due to the admissibility condition), the symmetries of the symbol are those of the tetrahedron, i.e.
\be
\left|
\begin{array}{ccc}
i&j&k\\
l&m&n
\end{array}\right|_q
=
\left|
\begin{array}{ccc}
j&i&k\\
m&l&n
\end{array}\right|_q
=
\left|
\begin{array}{ccc}
i&k&j\\
l&n&m
\end{array}\right|_q
=
\left|
\begin{array}{ccc}
i&m&n\\
l&j&k
\end{array}\right|_q
=
\left|
\begin{array}{ccc}
l&m&k\\
i&j&n
\end{array}\right|_q
=
\left|
\begin{array}{ccc}
l&j&n\\
i&m&k
\end{array}\right|_q.
\ee
In addition, the $6j$ symbol satisfies important relations which are at the heart of the topological invariance of the Turaev--Viro and the Ponzano--Regge models.
First, the Biedenharn--Elliot identity is given by
\be\label{BE}
\sum_i\omega_i^2
\left|
\begin{array}{ccc}
1&2&i\\
3&4&a
\end{array}\right|_q
\left|
\begin{array}{ccc}
5&6&i\\
3&4&b
\end{array}\right|_q
\left|
\begin{array}{ccc}
5&6&i\\
2&1&c
\end{array}\right|_q
=\left|
\begin{array}{ccc}
5&1&c\\
a&b&4
\end{array}\right|_q
\left|
\begin{array}{ccc}
c&2&6\\
3&b&a
\end{array}\right|_q,
\ee
where $1,2,\dots$ are shorthand notations for the spins $j_1,j_2,\dots$. It is immediate to see that this formula is closely related to 2--3 Pachner moves. Second, the $6j$ symbols satisfy the orthogonality relation given by
\be\label{ortho}
\sum_i\omega_a^2\omega_i^2
\left|
\begin{array}{ccc}
1&2&a\\
3&4&i
\end{array}\right|_q
\left|
\begin{array}{ccc}
1&2&b\\
3&4&i
\end{array}\right|_q=\delta_{a,b}.
\ee
It is also worth recalling that  we have for any spin $j$ the relation
\be\label{consistency2}
\omega^2=\omega_j^{-2}\sum_{k,l}\omega_k^2\omega_l^2 Y(j,k,l),
\ee
where, as specified with the symbol $Y(j,k,l)$, the sum is taken over the spins $k$ and $l$ which are such that the triple $(j,k,l)$ is admissible. This property explains the presence of the weights $\omega^{-2}$ in the Turaev--Viro partition function \eqref{TVsummand}.

Finally, let us mention for completeness the definition of the quantum $3jm$ symbol appearing in \eqref{triangle weight}. It is given by \cite{LB}
\be
\left(
\begin{array}{ccc}
j_1&j_2&j_3\\
m_1&m_2&-m_3
\end{array}\right)_q
=\f{(-1)^{j_1-j_2+m_3}}{\dim_q(j_3)^{1/2}}q^{(m_2-m_1)/6}\la j_1m_1j_2m_2|j_3m_3\ra_q.
\ee
The quantum Clebsch--Gordan coefficient appearing in this formula is given by \cite{N} (see references therein as well)
\ba
\la j_1m_1j_2m_2|j_3m_3\ra_q
&=&q^{\alpha/4}\Delta(j_1,j_2,j_3)\dim_q(j_3)^{1/2}\nonumber\\
&&\times\Big([j_1+m_1]_q![j_1-m_1]_q![j_2+m_2]_q![j_2-m_2]_q![j_3+m_3]_q![j_3-m_3]_q!\Big)^{1/2}\nonumber\\
&&\times\sum_z(-1)^z\f{\Big([j_1-m_1-z]_q![j_2+m_2-z]_q![j_1+j_2-j_3-z]_q!\Big)^{-1}}{q^{z(j_1+j_2+j_3+1)/2}[z]_q![z+j_3-j_2+m_1]_q![z+j_3-j_1-m_2]_q!},\qquad\qquad
\ea
where
\be
\alpha=j_1(j_1+1)+j_2(j_2+1)-j_3(j_3+1)+2(j_1j_2+j_1m_2-j_2m_1),
\ee
and where the sum runs over
\be
\max(-j_3+j_2-m_1,j_3+j_1+m_2)\leq z\leq\min(j_1-m_1,j_2+m_2,j_1+j_2-j_3).
\ee

\end{document}